\documentclass[twocolumn,prb,showpacs,preprintnumbers,amsmath,superscriptaddress,amssymb]{revtex4}

\usepackage{graphicx}
\usepackage{dcolumn}
\usepackage{bm}

\begin{document}

\title{Atomic multiplet calculation of $3d_{5/2} \rightarrow 4f$ resonant x-ray diffraction from Ho metal}

\author{M. W. Haverkort}
  \affiliation{ II. Physikalisches Institut, Universit{\"a}t zu K{\"o}ln, Z{\"u}lpicher Str. 77, D-50937 K{\"o}ln, Germany}
  \affiliation{Max Planck Institute for Solid State Research, Heisenbergstra{\ss}e 1, D-70569 Stuttgart Germany}
\author{C. Sch\"{u}{\ss}ler-Langeheine}
  \affiliation{ II. Physikalisches Institut, Universit{\"a}t zu K{\"o}ln, Z{\"u}lpicher Str. 77, D-50937 K{\"o}ln, Germany}
\author{C. F. Chang}
  \affiliation{ II. Physikalisches Institut, Universit{\"a}t zu K{\"o}ln, Z{\"u}lpicher Str. 77, D-50937 K{\"o}ln, Germany}
\author{M. Buchholz}
  \affiliation{ II. Physikalisches Institut, Universit{\"a}t zu K{\"o}ln, Z{\"u}lpicher Str. 77, D-50937 K{\"o}ln, Germany}
\author{H.-H. Wu}
  \affiliation{ II. Physikalisches Institut, Universit{\"a}t zu K{\"o}ln, Z{\"u}lpicher Str. 77, D-50937 K{\"o}ln, Germany}
\author{H. Ott}
  \affiliation{ II. Physikalisches Institut, Universit{\"a}t zu K{\"o}ln, Z{\"u}lpicher Str. 77, D-50937 K{\"o}ln, Germany}
\author{E. Schierle}
  \affiliation{Helmholtz-Zentrum Berlin für Materialien und Energie c/o BESSY, Albert-Einstein-Str. 15, D-12489 Berlin, Germany}
\author{D. Schmitz}
  \affiliation{Helmholtz-Zentrum Berlin für Materialien und Energie c/o BESSY, Albert-Einstein-Str. 15, D-12489 Berlin, Germany}
\author{A. Tanaka}
  \affiliation{Department of Quantum Matter, ADSM, Hiroshima University, Higashi-Hiroshima 739-8530, Japan}

\author{L. H. Tjeng}
  \affiliation{ II. Physikalisches Institut, Universit{\"a}t zu K{\"o}ln, Z{\"u}lpicher Str. 77, D-50937 K{\"o}ln, Germany}

\date{\today}

\begin{abstract}
We compare for Ho metal the x-ray absorption spectrum and the resonant soft x-ray diffraction spectra obtained at the $3d_{5/2} \rightarrow 4f$ ($M_5$) resonance for the magnetic 1st and 2nd order diffraction peaks $(0,0,\tau)$ and $(0,0,2\tau)$ with the result of an atomic multiplet calculation. We find a good agreement between experiment and simulation giving evidence that this kind of simulation is well suited to quantitatively analyze resonant soft x-ray diffraction data from correlated electron systems.

\end{abstract}

\maketitle
%75.25.+z 	Spin arrangements in magnetically ordered materials (including neutron and spin-polarized electron studies, synchrotron-source X-ray scattering, etc.)
%78.20.-e 	Optical properties of bulk materials and thin films
%78.70.Ck 	X-ray scattering
%78.70.Dm 	X-ray absorption spectra

Resonant soft x-ray diffraction (RSXD) at the transition metal $L_{2,3}$ ($2p \rightarrow 3d$) or lanthanide $M_{4,5}$ ($3d \rightarrow 4f$) resonance is a relative new tool to study orbital, charge or magnetic ordering phenomena.\cite{castleton:00a,abbamonte:02a,wilkins:03a,dhesi:04a,thomas:04a,abbamonte:04a,schuessler:05a,staub:05a,abbamonte:05a,huang:06a,rusydi:06a,staub:06a,scagnoli:06a} The technique makes use of the high sensitivity of an electronic transition at one of these resonances to the electronic and magnetic state of the scattering ions, which renders RSXD directly sensitive to super structures involving spatial variations of the electronic state.

In many interesting cases like transition-metal $L_{2,3}$ RSXD from transition-metal oxides and $M_{4,5}$ RSXD from lanthanide systems the resonance involves a strong interaction between the core-hole and the valence electrons making the resonant diffraction spectra strongly excitonic. A simple picture of RSXD in terms of the density of unoccupied states is hence not appropriate. In order to understand RSXD data from these systems quantitatively one needs a more realistic theoretical description. Fortunately, such a description has been developed for the closely related technique of x-ray-absorption spectroscopy (XAS) at transition metal $L_{2,3}$ edges and rare-earths $M_{4,5}$ edges.\cite{Gunnarsson83, Tanaka94, Groot94, Thole97} As RSXD and XAS involve the same optical transitions, i.e., the same orbitals are involved and the same transition matrix applies, one can expect that the theoretical treatment used for XAS can directly be applied to RSXD.

For the description of the excitonic intermediate state in RSXD, i.e., the final state in XAS, one needs to include the full electron-electron interaction. For transition-metal oxides this is usually done within a cluster calculations including the transition-metal ion and its oxygen ligands.\cite{Tanaka94, Groot94, Thole97} In lanthanide systems the overlap between the $4f$ wave function and its surrounding is usually so small\cite{Goedkoop88} that an atomic model (possibly including a crystal field) is sufficient. For XAS this technique has been used to obtain invaluable information on the local electronic structure of many transition metal and rare earth compounds. For RSXD, on the other hand, there are up to now only a few calculated spectra available in the literature.\cite{castleton:00a,thomas:04a,wilkins:05a,stojic:05a,schuessler:05a} Although these calculations match experimental data fairly well, they describe rather complex systems such that material properties and the applicability of the methods itself are probed at the same time. In order to verify that a description in analogy to XAS data is really appropriate for RSXD spectra a test with a simple model system is desirable. 

\begin{figure}[t]
    \includegraphics[clip,bb= 150 280 680 790,width=0.4\textwidth]{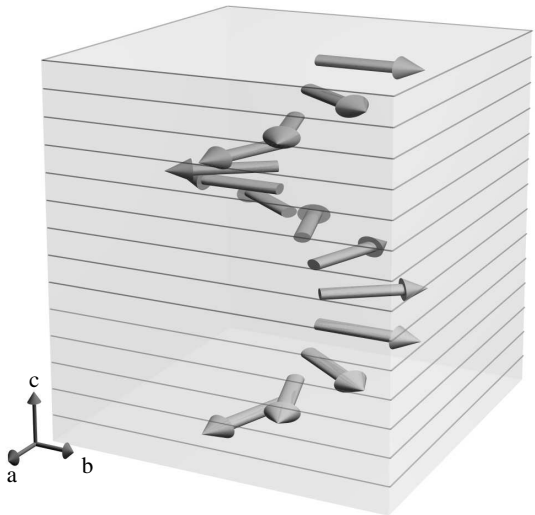}
    \caption{Schematic representation of the helical magnetic structure of holmium metal. The large arrows within the grey box symbolize the magnetization direction within the $ab$ planes.}
    \label{structure}
\end{figure}

\begin{figure}[t]
    \includegraphics[clip,bb= 80 370 515 730,width=0.4\textwidth]{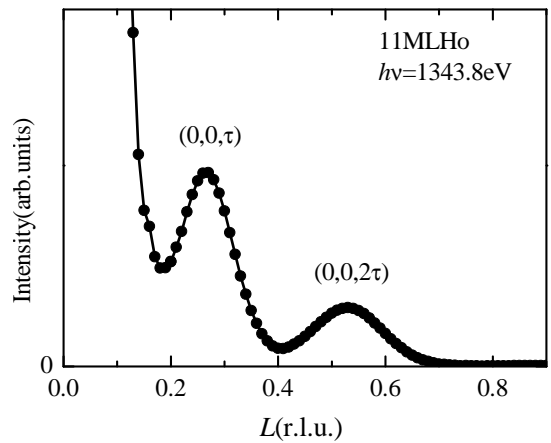}
    \caption{$L$-scan through the $(0,0,\tau)$ and $(0,0,2\tau)$ magnetic peaks at 26 K. The increasing intensity at low $L$-values is caused by specular reflectivity from the sample.}
    \label{lscan}
\end{figure}

In this paper we present a comparison of theoretical and experimental RSXD and XAS data from Ho metal at the $M_{5}$ threshold. Ho has a particularly simple and well established magnetic structure and has served as a reference system for the development of resonant magnetic scattering in the conventional \cite{gibbs:88a,hannon:88a} and soft x-ray range.\cite{schuessler:01a,ott:06a} Ho crystallizes in the hcp-structure; we refer in the following to the unit cell spanned by three orthogonal lattice vectors with $a$ = 3.58 {\AA}, $b$ = 6.20 {\AA}, and $c$ = 5.62 {\AA}. Below $T_N =$ 131.2 K bulk Ho orders with a helical magnetic structure (Fig.~\ref{structure}). The propagation vector of the helix $\vec{\tau}$ lies in the $c$ direction. The spin direction is perpendicular to $\vec{\tau}$.\cite{koehler:67a} Resonant scattering from this helical structure reveals two Bragg peaks (Fig.~\ref{lscan}), with one of them at $(0,0,\tau)$ corresponding to the total periodicity of the helix and the second one at $(0,0,2\tau)$ corresponding to half its periodicity.\cite{ott:06a} In the resonant magnetic scattering amplitude derived by Hannon et al.,\cite{hannon:88a} the $(0,0,\tau)$ peak is described by the term that is linear in the direction of the magnetic moment,$\vec{m}$, while the $(0,0,2\tau)$ peak relates to the term that is quadratic in $\vec{m}$.

At the Ho $M_5$ resonance the photon penetration depth changes strongly with energy. The corresponding change of the scattering volume leads to distortions of the resonance spectra when obtained from bulk samples\cite{spencer:05b}. For the experiment we therefore chose a 11-monolayer (ML) thin Ho film for which the probed volume is defined by the film thickness rather than by the photon-penetration depth. In the sample the Ho layer is sandwiched between Y layers with the consequence that the helix period length in this film of about 7.5 ML \cite{leiner:00a} is shorter than the bulk value of about 10 ML. The resulting larger $\tau$ value in the film allows for a better separation between magnetic peaks and specular reflectivity background at low momentum transfer, $q$. $T_N$ is reduced with respect to the bulk value to about 76 K.\cite{weschke:04a} While bulk Ho shows at low temperatures a lock-in transition to a commensurate magnetic structure and a canting of the spins out the $ab$ plane to a conical phase,\cite{koehler:67a} we observed none of these effects in our film.  

The diffraction experiments were carried out at the soft x-ray beamline UE46-PGM1 at the electron storage ring BESSY using the soft x-ray diffractometer designed at the FU Berlin. The experimental RSXD data were obtained by recording the scattered intensity at the respective peak maximum as a function of photon energy, keeping the momentum transfer constant. XAS data were taken from Ref.~\onlinecite{ott:06a} and were obtained from a 31 ML thick Ho film grown in \textit{situ} on a W(110) substrate. The cluster calculations have been done with the program XTLS8.3, \cite{Tanaka94} developed by A. Tanaka and previously used for the description of RSXD on La$_{1.8}$Sr$_{0.2}$NiO$_{4}$. \cite{schuessler:05a} The local model used for these calculations is the same as applied to describe the XAS on a series of Lanthanides by Thole \textit{et al.}. \cite{Thole85} It includes the full electron-electron repulsion, whereby we scaled the $4f-4f$ Slater integrals to 65\% and the $3d-4f$ Slater integrals to 75\% of the Hartree-Fock values. The Hartree-Fock values have been obtained with the use of Cowans program. \cite{Cowan81} The calculations have been done in spherical symmetry and on a single configuration. This is not a necessary restriction, as extensions to crystal and ligand field theory can easily be made. \cite{Gunnarsson83, Tanaka94, Groot94, Thole97} For Ho metal the crystal-fields acting on the $4f$ shell are small, though, and therefore spherical symmetry is sufficient.

XAS and RSXD can be most easily formalized if one starts from linear response theory. For each atom one can define a scattering tensor $F_{j}$ which describes the relation between the components of the scattered radial wave $E_{out}^j$ at a distance $R$ of the scatterer and those of the incoming plane wave $E_{in}^{j}$ at frequency $\omega$. $E_{out}^j(\omega)=F_{j}(\omega)\cdot E_{in}^j(\omega) (e^{\imath \omega R /c}/R)$. The scattering tensor at a specific scattering vector $\vec{q}$ can be written as a sum over the scattering tensors of each atom multiplied by the appropriate phase: $F(\vec{q},\omega)=\sum_n e^{i \vec{q}\cdot\vec{R}_n}\sum_j e^{i \vec{q}\cdot \vec{r}_j} F_{j}(\omega)$, where the first sum runs over all unit cells at positions $\vec{R}_n$ in the probed volume and the second sum over the ions at positions $\vec{r}_j$ in one unit cell. If one now defines $\varepsilon_{in}$ ($\varepsilon_{out}$) as the polarization of the incoming (outgoing) photons one can write the intensities of RSXD and XAS as:
\begin{eqnarray}
  I_{RSXD}(\vec{q}, \omega) &=& \left| \varepsilon_{out}^{*} \cdot F(\vec{q}, \omega) \cdot \varepsilon_{in} \right|^2 \\
  I_{XAS}(\omega) &=& -\frac{4 \pi c}{\omega}\Im\left[ \varepsilon^{*}_{in} \cdot F(\vec{q}=0,\omega) \cdot \varepsilon_{in} \right]
\end{eqnarray}
The relation between the absorption and scattered intensity is given by the optical theorem, which follows from conservation of energy. \cite{Newton76} The real and imaginary part of $F(\vec{q},\omega)$ are related by the Kramers-Kronig transformations in the $\omega$ domain. 

The scattered intensity depends on the polarization of the incoming and outgoing photons; hence it is a $3\times 3$ tensor, whereas the absorption depends only on the incoming polarization and is described by the diagonal tensor elements only. By rotating the polarization vector (real as well as complex rotations) one can write all components of the $F$ tensor as a linear combination of diagonal terms. For example, defining $(x+y)$ as the light polarized in the $x+y$ direction and $(x+\imath y)$ as right circular polarized light, we can describe the scattering of $x$-polarized incoming to $y$-polarized outgoing light by:
\begin{equation}
F_{xy}=F_{(x+y)(x+y)}+\imath F_{(x+\imath y)(x+\imath y)}-\frac{1+\imath}{2}(F_{xx}+F_{yy})
\end{equation}
This gives a direct relation between XAS spectra recorded with different incoming light polarizations and all possible RSXD channels.

\begin{figure}[t]
    \includegraphics[width=0.45\textwidth]{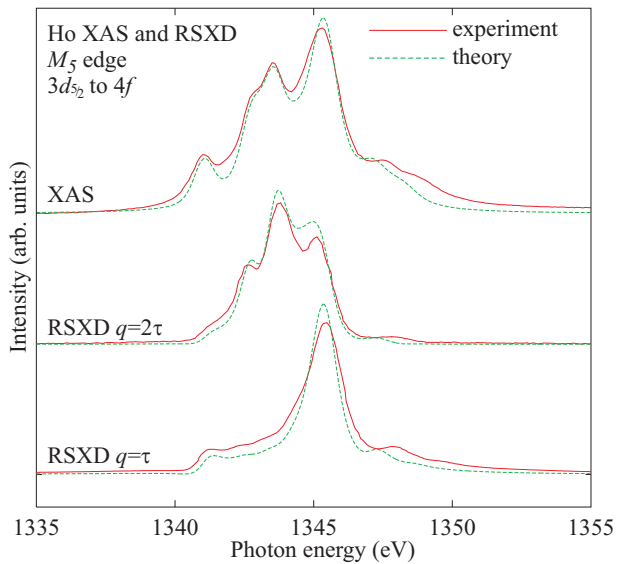}
    \caption{(color online) Experimental and theoretical Ho $M_{5}$ XAS and RSXD spectra.}
    \label{spectra}
\end{figure}

In Fig.~\ref{spectra} we show the theoretical and experimental XAS and RSXD spectra of the Ho $M_5$ edge. Although the probed resonance is the same for all three experiments, one can clearly see that each measurement, XAS, RSXD at $\vec{q}=\vec{\tau}$, and RSXD at $\vec{q}=2\vec{\tau}$ has a very different spectral shape. The theoretical calculations reproduce these different spectra quite well. The agreement between theory and RSXD is equally good as the agreement between theory and XAS. This is pleasing as it is known that cluster calculations can reproduce XAS for many transition metal and rare-earth compounds rather well. Obviously the level of agreement is the same for our RSXD spectra.

The huge differences in spectral shape between XAS and RSXD at different $\vec{q}$ values can be understood if one realizes which different elements of the $F(\vec{q},\omega)$ tensor are probed by the different signals. We chose a notation with respect to an orthogonal coordinate system aligned with the crystalline axes. For a magnetic ion with the magnetization in the $x$ direction in otherwise spherical symmetry $F_{j}(\omega)$ can be written as:
\begin{equation}
F_{j}=\left(
  \begin{array}{ccc}
    F_{xx} & F_{yx} & F_{zx} \\
    F_{xy} & F_{yy} & F_{zy} \\
    F_{xz} & F_{yz} & F_{zz} \\
  \end{array}
\right)=\left(
  \begin{array}{ccc}
    F_{\parallel} & 0 & 0 \\
    0 & F_{\perp} & \imath F_{\odot} \\
    0 & -\imath F_{\odot} & F_{\perp} \\
  \end{array}
\right)
\end{equation}
The diagonal elements of $F_{j}(\omega)$ refer to scattering processes which do not alter the polarization of the scattered light. The off-diagonal elements describe scattering processes involving changes of the light polarization. Magnetic scattering breaks time-reversal symmetry and is hence (on a basis with one of the principle axes parallel to the magnetization, in the present case $x$) described by an antisymmetric off-diagonal part of $F_{j}(\omega)$. 

It is tempting to relate the symmetric part of the $F$ tensor to charge scattering and anti-symmetric part to magnetic scattering. Care should be taken, however, as a local magnetization also produces symmetric contributions to the $F$ tensor. One reason is that a local magnetization can induce an orbital momentum, which is related to a non-spherical charge distribution and therefore gives rise to charge scattering. But there is also an effect creating symmetric contributions to the $F$ tensor due to a local magnetization that is not related to a local charge distribution, but pure magnetically: At $L_{2,3}$ or $M_{4,5}$ edges the core hole is created in a shell with orbital momentum ($2p$ or $3d$) and large spin-orbit coupling. The magnetization of the valence electrons interacts via the multipolar part of the Coulomb interaction (direct and exchange) with the core electrons. This results in a shift of resonance energy between different polarizations, reflected in symmetric and antisymmetric contributions to the $F$ tensor. This second contribution would for example also be present in $d^5$ or $f^7$ configurations. 

In order to analyze the different contributions to the different spectra in Fig. 1 we use the simplest sufficient model, which is reproducing the two observed Bragg-peaks. It consists of a chain of four atoms labeled $A$,$B$,$C$, and $D$ stacked in the $z$ direction with the magnetic moment alternating in the $x$, $y$, $-x$, $-y$ direction. In our model we have four different scattering tensors. $F_B$ is related to $F_A$ by a $90^\circ$ rotation around $z$ and so on.
\begin{equation}
\begin{array}{ll}
  F_{A}=\left(
        \begin{array}{ccc}
          F_{\parallel} & 0 & 0 \\
          0 & F_{\perp} & \imath F_{\odot} \\
          0 & -\imath F_{\odot} & F_{\perp} \\
        \end{array}
      \right) & F_{C}=\left(
        \begin{array}{ccc}
          F_{\parallel} & 0 & 0 \\
          0 & F_{\perp} & -\imath F_{\odot} \\
          0 & \imath F_{\odot} & F_{\perp} \\
        \end{array}
      \right) \\
      &\\
  F_{B}=\left(
        \begin{array}{ccc}
          F_{\perp} & 0 & -\imath F_{\odot} \\
          0 & F_{\parallel} & 0 \\
          \imath F_{\odot} & 0 & F_{\perp} \\
        \end{array}
      \right) & F_{D}=\left(
        \begin{array}{ccc}
          F_{\perp} & 0 & \imath F_{\odot} \\
          0 & F_{\parallel} & 0 \\
          -\imath F_{\odot} & 0 & F_{\perp} \\
        \end{array}
      \right)
\end{array}
\end{equation}
The generalization to helices with longer periods or incommensurate order is straightforward. The neglect of crystal field effects and assumption of a spherical symmetry are no necessary simplifications, but for Ho turns out to be sufficient, most probably because crystal field effects are smaller than spin-orbit coupling and magnetic interactions.

The $F$ tensor corresponding to a phase repetition after four lattice spacings [$\vec{q}=(0,0,\tau)$] then becomes:
\begin{eqnarray}
\nonumber F[\vec{q}&=&(0,0,\tau),\omega]=F_{A}+\imath F_{B} -F_{C}-\imath F_{D}\\
         &=&\left(
                  \begin{array}{ccc}
                    0 & 0 & 2 F_{\odot} \\
                    0 & 0 & 2 \imath F_{\odot} \\
                    -2 F_{\odot} & - \imath 2 F_{\odot} & 0 \\
                  \end{array}
            \right)
\end{eqnarray}
The $F$ tensor corresponding to a phase repetition after two lattice spacings [$\vec{q}=(0,0,2 \tau)$] is:
\begin{eqnarray}
\nonumber F[\vec{q}&=&(0,0,2 \tau),\omega]=F_{A}- F_{B} +F_{C}-F_{D}\\
         &=&\left(
                  \begin{array}{ccc}
                    2(F_{\parallel}-F_{\perp}) & 0 & 0 \\
                    0 & 2(F_{\perp}-F_{\parallel}) & 0 \\
                    0 & 0 & 0 \\
                  \end{array}
            \right)
\end{eqnarray}
The $(0,0,\tau)$ peak is hence only given by the off-diagonal element of the single-scatterer's tensor, $F_{\odot}$, and the $(0,0,2\tau)$ peak only by the differences between the diagonal terms $F_{\perp} - F_{\parallel}$. This property is also conserved for other helix periodicities. 

The $F$ tensor for XAS [$\vec{q}=(0,0,0)$] becomes:
\begin{eqnarray}
\nonumber F[\vec{q}&=&(0,0,0),\omega]=F_{A}+F_{B}+F_{C}+F_{D}\\
         &=&\left(
                  \begin{array}{ccc}
                    2(F_{\parallel}+F_{\perp}) & 0 & 0 \\
                    0 & 2(F_{\perp}+F_{\parallel}) & 0 \\
                    0 & 0 & 4 F_{\perp} \\
                  \end{array}
            \right)
\end{eqnarray}
In our case the XAS measurement have been preformed in the paramagnetic phase above the N\'eel temperature which means that the local spin-direction should be averaged over all possible directions and the $F$ tensor for this case becomes:

\begin{eqnarray}
\nonumber F[\vec{q}&=&(0,0,0],\omega)=\\
         &=&4/3\left(
                  \begin{array}{ccc}
                    F_{\parallel}+2F_{\perp} & 0 & 0 \\
                    0 & F_{\parallel}+2F_{\perp} & 0 \\
                    0 & 0 & F_{\parallel}+2F_{\perp} \\
                  \end{array}
            \right)
\end{eqnarray}

\begin{figure}[t]
    \includegraphics[width=0.4\textwidth]{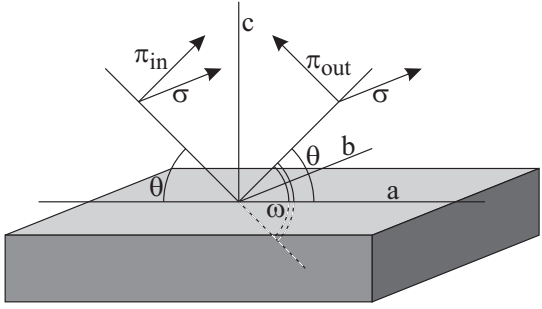}
    \caption{Experimental geometry. The direction of the Poyinting vectors are determined by the Bragg condition. The polarizations can be defined as: $\sigma=(0,1,0)$, $\pi_{in}=(\sin \theta,0,\cos \theta)$, and $\pi_{out}=(-\sin \theta,0,\cos \theta)$.}
    \label{geometry}
\end{figure}

We have so far discussed the scattering tensors within the reference frame of the crystal, which makes the mathematical formalism a bit more transparent. Usually one writes the polarization vectors with respect to the directions parallel to the scattering plane ($\pi$) and perpendicular to it ($\sigma$). For Ho with the scattering vector in the $z$ direction and if we choose $a$ to be in the scattering plane we find the polarizations to be equal to:
\begin{eqnarray}
\sigma_{in} = \sigma_{out} &=&(0,1,0)\\
\pi_{in}&=&(\sin \theta,0,\cos \theta)\\
\pi_{out}&=&(-\sin \theta,0,\cos \theta)
\end{eqnarray}
with $\theta = \omega/2$ and $\omega \approx 25^\circ$ for $(0,0,\tau)$ and $\omega \approx 51.4^\circ$ for $(0,0,2\tau)$ in the considered energy range. With the scattering angles as defined in Fig. 4.
One now has all ingredients to straightforwardly calculate the RSXD and XAS intensity. By substituting the $F$ tensor as given in formula 6,7, and 9 into equation 1 or 2 for RSXD and XAS respectively one finds the intensity with the polarizations as given in formulas 10-12. Note that since the $F$ tensors as written in equation 6, 7, and 9 depend only on a single function, the spectral lineshape does not depend on the polarization or on the rotation of the sample around $\vec{q}$. The total intensity still does, though.

The analyzes of the scattering intensity and line-shape as presented here can be directly related to the results derived by Hannon \textit{et al.} \cite{hannon:88a}, if one changes the basis of the $F$ tensor from linear polarized light ($x$,$y$,$z$) to circular polarized light ($-1=1/\sqrt{2}(x-\imath y)$,$0=z$,$1=1/\sqrt{2}(-x-\imath y)$). For a site with a magnetization in the $z$ direction one finds that $1/2(F^{j}_{1}+F^{j}_{-1})=1/2(F^{j}_{xx}+F^{j}_{yy})=F^{j}_{\perp}$, $1/2(F^{j}_{-1}-F^{j}_{1})=1/2\imath(F^{j}_{xy}-F^{j}_{yx})=F^{j}_{\odot}$, and $F_{0}=F_{\parallel}$. The formulas of Hannon \textit{et al.} for dipole interactions are retrieved if one uses this basis and realizes that in spherical symmetry the $F$ tensor for a magnetization in arbitrary direction is given by a simple rotation of the $F$ tensor magnetized in the $z$ direction; $F_j(\theta,\phi)=R_{\theta,\phi} \cdot F_{j}(z) \cdot R_{\theta,\phi}^{\dagger}$. Whereby $F_j(z)$ is the local $F$ tensor for a magnetization in the $z$ direction, $F_{j}(\theta,\phi)$ the local $F$ tensor for a magnetization in the $\theta, \phi$ direction and $R_{\theta,\phi}$ a rotation matrix rotating a vector in the $z$ direction to a vector in the $\theta, \phi$ direction.

To summarize, we have shown that it is straightforwardly possible to simulate RSXD spectra from a $4f$ system using configuration interaction cluster calculations with very good agreement between experiment and simulation. This technique is well developed for XAS and can without further modification directly be used for RSXD spectra. With the use of such analyzes one can expect to obtain a wealth of spectroscopic information from RSXD on the ordered electronic structure like orbital, spin and charge ordering quantitatively.

This work has funded by the DFG through SFB 608. Experiments at BESSY were supported by the BMBF through project 05 ES3XBA/5. We appreciate experimental support by the BESSY and HMI staff. We thank E. Weschke for the permission to use his soft x-ray diffractometer and L. Hamdan for skillful technical assistance. The 11-ML thin Ho film was grown in the group of H. Zabel at the Ruhr-University Bochum.


\begin{thebibliography}{31}
\expandafter\ifx\csname natexlab\endcsname\relax\def\natexlab#1{#1}\fi
\expandafter\ifx\csname bibnamefont\endcsname\relax
  \def\bibnamefont#1{#1}\fi
\expandafter\ifx\csname bibfnamefont\endcsname\relax
  \def\bibfnamefont#1{#1}\fi
\expandafter\ifx\csname citenamefont\endcsname\relax
  \def\citenamefont#1{#1}\fi
\expandafter\ifx\csname url\endcsname\relax
  \def\url#1{\texttt{#1}}\fi
\expandafter\ifx\csname urlprefix\endcsname\relax\def\urlprefix{URL }\fi
\providecommand{\bibinfo}[2]{#2}
\providecommand{\eprint}[2][]{\url{#2}}

\bibitem[{\citenamefont{Castleton and Altarelli}(2000)}]{castleton:00a}
\bibinfo{author}{\bibfnamefont{C.~W.~M.} \bibnamefont{Castleton}}
  \bibnamefont{and}
  \bibinfo{author}{\bibfnamefont{M.}~\bibnamefont{Altarelli}},
  \bibinfo{journal}{Phys.~Rev.~B} \textbf{\bibinfo{volume}{62}},
  \bibinfo{pages}{1033} (\bibinfo{year}{2000}).

\bibitem[{\citenamefont{Abbamonte et~al.}(2002)\citenamefont{Abbamonte, Venema,
  Rusydi, Sawatzky, Logvenov, and Bozovic}}]{abbamonte:02a}
\bibinfo{author}{\bibfnamefont{P.}~\bibnamefont{Abbamonte}},
  \bibinfo{author}{\bibfnamefont{L.}~\bibnamefont{Venema}},
  \bibinfo{author}{\bibfnamefont{A.}~\bibnamefont{Rusydi}},
  \bibinfo{author}{\bibfnamefont{G.~A.} \bibnamefont{Sawatzky}},
  \bibinfo{author}{\bibfnamefont{G.}~\bibnamefont{Logvenov}}, \bibnamefont{and}
  \bibinfo{author}{\bibfnamefont{I.}~\bibnamefont{Bozovic}},
  \bibinfo{journal}{Science} \textbf{\bibinfo{volume}{297}},
  \bibinfo{pages}{581} (\bibinfo{year}{2002}).

\bibitem[{\citenamefont{Wilkins et~al.}(2003)\citenamefont{Wilkins, Spencer,
  Hatton, Collins, Roper, Prabhakaran, and Boothroyd}}]{wilkins:03a}
\bibinfo{author}{\bibfnamefont{S.~B.} \bibnamefont{Wilkins}},
  \bibinfo{author}{\bibfnamefont{P.~D.} \bibnamefont{Spencer}},
  \bibinfo{author}{\bibfnamefont{P.~D.} \bibnamefont{Hatton}},
  \bibinfo{author}{\bibfnamefont{S.~P.} \bibnamefont{Collins}},
  \bibinfo{author}{\bibfnamefont{M.~D.} \bibnamefont{Roper}},
  \bibinfo{author}{\bibfnamefont{D.}~\bibnamefont{Prabhakaran}},
  \bibnamefont{and} \bibinfo{author}{\bibfnamefont{A.~T.}
  \bibnamefont{Boothroyd}}, \bibinfo{journal}{Phys.~Rev.~Lett.}
  \textbf{\bibinfo{volume}{91}}, \bibinfo{pages}{167205}
  (\bibinfo{year}{2003}).

\bibitem[{\citenamefont{Dhesi et~al.}(2004)\citenamefont{Dhesi, Mirone, {De
  Nadai}, Ohresser, Bencok, Brookes, Reutler, Revcolevschi, Tagliaferri,
  Toulemonde et~al.}}]{dhesi:04a}
\bibinfo{author}{\bibfnamefont{S.~S.} \bibnamefont{Dhesi}},
  \bibinfo{author}{\bibfnamefont{A.}~\bibnamefont{Mirone}},
  \bibinfo{author}{\bibfnamefont{C.}~\bibnamefont{{De Nadai}}},
  \bibinfo{author}{\bibfnamefont{P.}~\bibnamefont{Ohresser}},
  \bibinfo{author}{\bibfnamefont{P.}~\bibnamefont{Bencok}},
  \bibinfo{author}{\bibfnamefont{N.~B.} \bibnamefont{Brookes}},
  \bibinfo{author}{\bibfnamefont{P.}~\bibnamefont{Reutler}},
  \bibinfo{author}{\bibfnamefont{A.}~\bibnamefont{Revcolevschi}},
  \bibinfo{author}{\bibfnamefont{A.}~\bibnamefont{Tagliaferri}},
  \bibinfo{author}{\bibfnamefont{O.}~\bibnamefont{Toulemonde}},
  \bibnamefont{et~al.}, \bibinfo{journal}{Phys.~Rev.~Lett.}
  \textbf{\bibinfo{volume}{92}}, \bibinfo{pages}{056403}
  (\bibinfo{year}{2004}).

\bibitem[{\citenamefont{Thomas et~al.}(2004)\citenamefont{Thomas, Hill,
  Grenier, Kim, Abbamonte, Venema, Rusydi, Tomioka, Tokura, McMorrow
  et~al.}}]{thomas:04a}
\bibinfo{author}{\bibfnamefont{K.~J.} \bibnamefont{Thomas}},
  \bibinfo{author}{\bibfnamefont{J.~P.} \bibnamefont{Hill}},
  \bibinfo{author}{\bibfnamefont{S.}~\bibnamefont{Grenier}},
  \bibinfo{author}{\bibfnamefont{Y.-J.} \bibnamefont{Kim}},
  \bibinfo{author}{\bibfnamefont{P.}~\bibnamefont{Abbamonte}},
  \bibinfo{author}{\bibfnamefont{L.}~\bibnamefont{Venema}},
  \bibinfo{author}{\bibfnamefont{A.}~\bibnamefont{Rusydi}},
  \bibinfo{author}{\bibfnamefont{Y.}~\bibnamefont{Tomioka}},
  \bibinfo{author}{\bibfnamefont{Y.}~\bibnamefont{Tokura}},
  \bibinfo{author}{\bibfnamefont{D.~F.} \bibnamefont{McMorrow}},
  \bibnamefont{et~al.}, \bibinfo{journal}{Phys.~Rev.~Lett.}
  \textbf{\bibinfo{volume}{92}}, \bibinfo{pages}{237204}
  (\bibinfo{year}{2004}).

\bibitem[{\citenamefont{Abbamonte et~al.}(2004)\citenamefont{Abbamonte,
  Blumberg, Rusydi, Gozar, Evans, Siegrist, Venema, Eisaki, Isaacs, and
  Sawatzky}}]{abbamonte:04a}
\bibinfo{author}{\bibfnamefont{P.}~\bibnamefont{Abbamonte}},
  \bibinfo{author}{\bibfnamefont{G.}~\bibnamefont{Blumberg}},
  \bibinfo{author}{\bibfnamefont{A.}~\bibnamefont{Rusydi}},
  \bibinfo{author}{\bibfnamefont{A.}~\bibnamefont{Gozar}},
  \bibinfo{author}{\bibfnamefont{P.~G.} \bibnamefont{Evans}},
  \bibinfo{author}{\bibfnamefont{T.}~\bibnamefont{Siegrist}},
  \bibinfo{author}{\bibfnamefont{L.}~\bibnamefont{Venema}},
  \bibinfo{author}{\bibfnamefont{H.}~\bibnamefont{Eisaki}},
  \bibinfo{author}{\bibfnamefont{E.~D.} \bibnamefont{Isaacs}},
  \bibnamefont{and} \bibinfo{author}{\bibfnamefont{G.~A.}
  \bibnamefont{Sawatzky}}, \bibinfo{journal}{Nature (London)}
  \textbf{\bibinfo{volume}{431}}, \bibinfo{pages}{1078} (\bibinfo{year}{2004}).

\bibitem[{\citenamefont{Sch\"u{\ss}ler-Langeheine
  et~al.}(2005)\citenamefont{Sch\"u{\ss}ler-Langeheine, Schlappa, Tanaka, Hu,
  Chang, Schierle, Benomar, Ott, Weschke, Kaindl et~al.}}]{schuessler:05a}
\bibinfo{author}{\bibfnamefont{C.}~\bibnamefont{Sch\"u{\ss}ler-Langeheine}},
  \bibinfo{author}{\bibfnamefont{J.}~\bibnamefont{Schlappa}},
  \bibinfo{author}{\bibfnamefont{A.}~\bibnamefont{Tanaka}},
  \bibinfo{author}{\bibfnamefont{Z.}~\bibnamefont{Hu}},
  \bibinfo{author}{\bibfnamefont{C.~F.} \bibnamefont{Chang}},
  \bibinfo{author}{\bibfnamefont{E.}~\bibnamefont{Schierle}},
  \bibinfo{author}{\bibfnamefont{M.}~\bibnamefont{Benomar}},
  \bibinfo{author}{\bibfnamefont{H.}~\bibnamefont{Ott}},
  \bibinfo{author}{\bibfnamefont{E.}~\bibnamefont{Weschke}},
  \bibinfo{author}{\bibfnamefont{G.}~\bibnamefont{Kaindl}},
  \bibnamefont{et~al.}, \bibinfo{journal}{Phys.~Rev.~Lett.}
  \textbf{\bibinfo{volume}{95}}, \bibinfo{pages}{156402}
  (\bibinfo{year}{2005}).

\bibitem[{\citenamefont{Staub et~al.}(2005)\citenamefont{Staub, Scagnoli,
  Mulders, Katsumata, Honda, Grimmer, Horisberger, and Tonnerre}}]{staub:05a}
\bibinfo{author}{\bibfnamefont{U.}~\bibnamefont{Staub}},
  \bibinfo{author}{\bibfnamefont{V.}~\bibnamefont{Scagnoli}},
  \bibinfo{author}{\bibfnamefont{A.~M.} \bibnamefont{Mulders}},
  \bibinfo{author}{\bibfnamefont{K.}~\bibnamefont{Katsumata}},
  \bibinfo{author}{\bibfnamefont{Z.}~\bibnamefont{Honda}},
  \bibinfo{author}{\bibfnamefont{H.}~\bibnamefont{Grimmer}},
  \bibinfo{author}{\bibfnamefont{M.}~\bibnamefont{Horisberger}},
  \bibnamefont{and} \bibinfo{author}{\bibfnamefont{J.~M.}
  \bibnamefont{Tonnerre}}, \bibinfo{journal}{Phys.~Rev.~B}
  \textbf{\bibinfo{volume}{71}}, \bibinfo{pages}{214421}
  (\bibinfo{year}{2005}).

\bibitem[{\citenamefont{Abbamonte et~al.}(2005)\citenamefont{Abbamonte, Rusydi,
  Smadici, Gu, Sawatzky, and Feng}}]{abbamonte:05a}
\bibinfo{author}{\bibfnamefont{P.}~\bibnamefont{Abbamonte}},
  \bibinfo{author}{\bibfnamefont{A.}~\bibnamefont{Rusydi}},
  \bibinfo{author}{\bibfnamefont{S.}~\bibnamefont{Smadici}},
  \bibinfo{author}{\bibfnamefont{G.~D.} \bibnamefont{Gu}},
  \bibinfo{author}{\bibfnamefont{G.~A.} \bibnamefont{Sawatzky}},
  \bibnamefont{and} \bibinfo{author}{\bibfnamefont{D.~L.} \bibnamefont{Feng}},
  \bibinfo{journal}{Nature Physics} \textbf{\bibinfo{volume}{1}},
  \bibinfo{pages}{155} (\bibinfo{year}{2005}).

\bibitem[{\citenamefont{Huang et~al.}(2006)\citenamefont{Huang, Lin, Okamoto,
  Chao, Jeng, Guo, Hsu, Huang, Ling, Wu et~al.}}]{huang:06a}
\bibinfo{author}{\bibfnamefont{D.~J.} \bibnamefont{Huang}},
  \bibinfo{author}{\bibfnamefont{H.-J.} \bibnamefont{Lin}},
  \bibinfo{author}{\bibfnamefont{J.}~\bibnamefont{Okamoto}},
  \bibinfo{author}{\bibfnamefont{K.~S.} \bibnamefont{Chao}},
  \bibinfo{author}{\bibfnamefont{H.-T.} \bibnamefont{Jeng}},
  \bibinfo{author}{\bibfnamefont{G.~Y.} \bibnamefont{Guo}},
  \bibinfo{author}{\bibfnamefont{C.-H.} \bibnamefont{Hsu}},
  \bibinfo{author}{\bibfnamefont{C.-M.} \bibnamefont{Huang}},
  \bibinfo{author}{\bibfnamefont{D.~C.} \bibnamefont{Ling}},
  \bibinfo{author}{\bibfnamefont{W.~B.} \bibnamefont{Wu}},
  \bibnamefont{et~al.}, \bibinfo{journal}{Phys.~Rev.~Lett.}
  \textbf{\bibinfo{volume}{96}}, \bibinfo{pages}{096401}
  (\bibinfo{year}{2006}).

\bibitem[{\citenamefont{Rusydi et~al.}(2006)\citenamefont{Rusydi, Abbamonte,
  Eisaki, Fujimaki, Blumberg, Uchida, and Sawatzky}}]{rusydi:06a}
\bibinfo{author}{\bibfnamefont{A.}~\bibnamefont{Rusydi}},
  \bibinfo{author}{\bibfnamefont{P.}~\bibnamefont{Abbamonte}},
  \bibinfo{author}{\bibfnamefont{H.}~\bibnamefont{Eisaki}},
  \bibinfo{author}{\bibfnamefont{Y.}~\bibnamefont{Fujimaki}},
  \bibinfo{author}{\bibfnamefont{G.}~\bibnamefont{Blumberg}},
  \bibinfo{author}{\bibfnamefont{S.}~\bibnamefont{Uchida}}, \bibnamefont{and}
  \bibinfo{author}{\bibfnamefont{G.~A.} \bibnamefont{Sawatzky}},
  \bibinfo{journal}{Phys.~Rev.~Lett.} \textbf{\bibinfo{volume}{97}},
  \bibinfo{pages}{016403} (\bibinfo{year}{2006}).

\bibitem[{\citenamefont{Staub et~al.}(2006)\citenamefont{Staub, Scagnoli,
  Mulders, Janousch, Honda, and Tonnere}}]{staub:06a}
\bibinfo{author}{\bibfnamefont{U.}~\bibnamefont{Staub}},
  \bibinfo{author}{\bibfnamefont{V.}~\bibnamefont{Scagnoli}},
  \bibinfo{author}{\bibfnamefont{A.~M.} \bibnamefont{Mulders}},
  \bibinfo{author}{\bibfnamefont{M.}~\bibnamefont{Janousch}},
  \bibinfo{author}{\bibfnamefont{Z.}~\bibnamefont{Honda}}, \bibnamefont{and}
  \bibinfo{author}{\bibfnamefont{J.~M.} \bibnamefont{Tonnere}},
  \bibinfo{journal}{Europhys. Lett.} \textbf{\bibinfo{volume}{76}},
  \bibinfo{pages}{926} (\bibinfo{year}{2006}).

\bibitem[{\citenamefont{Scagnoli et~al.}(2006)\citenamefont{Scagnoli, Staub,
  Mulders, Janousch, Meijer, Hammerl, Tonnerre, and Stojic}}]{scagnoli:06a}
\bibinfo{author}{\bibfnamefont{V.}~\bibnamefont{Scagnoli}},
  \bibinfo{author}{\bibfnamefont{U.}~\bibnamefont{Staub}},
  \bibinfo{author}{\bibfnamefont{A.~M.} \bibnamefont{Mulders}},
  \bibinfo{author}{\bibfnamefont{M.}~\bibnamefont{Janousch}},
  \bibinfo{author}{\bibfnamefont{G.~I.} \bibnamefont{Meijer}},
  \bibinfo{author}{\bibfnamefont{G.}~\bibnamefont{Hammerl}},
  \bibinfo{author}{\bibfnamefont{J.~M.} \bibnamefont{Tonnerre}},
  \bibnamefont{and} \bibinfo{author}{\bibfnamefont{N.}~\bibnamefont{Stojic}},
  \bibinfo{journal}{Phys.~Rev.~B} \textbf{\bibinfo{volume}{73}},
  \bibinfo{pages}{100409} (\bibinfo{year}{2006}).

\bibitem[{\citenamefont{Gunnarsson et~al.}(1983)\citenamefont{Gunnarsson,
  Sch\"onhammer, Fuggle, Hillebrecht, Esteva, Karnatak, and
  Hillebrand}}]{Gunnarsson83}
\bibinfo{author}{\bibfnamefont{O.}~\bibnamefont{Gunnarsson}},
  \bibinfo{author}{\bibfnamefont{K.}~\bibnamefont{Sch\"onhammer}},
  \bibinfo{author}{\bibfnamefont{J.~C.} \bibnamefont{Fuggle}},
  \bibinfo{author}{\bibfnamefont{F.~U.} \bibnamefont{Hillebrecht}},
  \bibinfo{author}{\bibfnamefont{J.~M.} \bibnamefont{Esteva}},
  \bibinfo{author}{\bibfnamefont{R.~C.} \bibnamefont{Karnatak}},
  \bibnamefont{and}
  \bibinfo{author}{\bibfnamefont{B.}~\bibnamefont{Hillebrand}},
  \bibinfo{journal}{Phys. Rev. B} \textbf{\bibinfo{volume}{28}},
  \bibinfo{pages}{7330} (\bibinfo{year}{1983}).

\bibitem[{Tan()}]{Tanaka94}
\bibinfo{note}{A. Tanaka and T. Jo, J. Phys. Soc. Jpn. \textbf{63}, 2788
  (1994).}

\bibitem[{Gro()}]{Groot94}
\bibinfo{note}{F. M. F. de Groot, J. Electron Spectrosc. Relat. Phenom.
  \textbf{67}, 529 (1994).}

\bibitem[{Tho()}]{Thole97}
\bibinfo{note}{Theo Thole Memorial Issue, edited by A. P. Hitchcock, G. E.
  McGuire, and J. J. Pireaux [J. Electron Spectrosc. Relat. Phenom. 86, 1
  (1997)].}

\bibitem[{\citenamefont{Goedkoop et~al.}(1988)\citenamefont{Goedkoop, Thole,
  van~der Laan, Sawatzky, de~Groot, and Fuggle}}]{Goedkoop88}
\bibinfo{author}{\bibfnamefont{J.~B.} \bibnamefont{Goedkoop}},
  \bibinfo{author}{\bibfnamefont{B.~T.} \bibnamefont{Thole}},
  \bibinfo{author}{\bibfnamefont{G.}~\bibnamefont{van~der Laan}},
  \bibinfo{author}{\bibfnamefont{G.~A.} \bibnamefont{Sawatzky}},
  \bibinfo{author}{\bibfnamefont{F.~M.~F.} \bibnamefont{de~Groot}},
  \bibnamefont{and} \bibinfo{author}{\bibfnamefont{J.~C.}
  \bibnamefont{Fuggle}}, \bibinfo{journal}{Phys. Rev. B}
  \textbf{\bibinfo{volume}{37}}, \bibinfo{pages}{2086} (\bibinfo{year}{1988}).

\bibitem[{\citenamefont{Wilkins et~al.}(2005)\citenamefont{Wilkins, Stoji\'{c},
  Beale, Binggeli, Castleton, Bencok, Prabhakaran, Boothroyd, Hatton, and
  Altarelli}}]{wilkins:05a}
\bibinfo{author}{\bibfnamefont{S.~B.} \bibnamefont{Wilkins}},
  \bibinfo{author}{\bibfnamefont{N.}~\bibnamefont{Stoji\'{c}}},
  \bibinfo{author}{\bibfnamefont{T.~A.~W.} \bibnamefont{Beale}},
  \bibinfo{author}{\bibfnamefont{N.}~\bibnamefont{Binggeli}},
  \bibinfo{author}{\bibfnamefont{C.~W.~M.} \bibnamefont{Castleton}},
  \bibinfo{author}{\bibfnamefont{P.}~\bibnamefont{Bencok}},
  \bibinfo{author}{\bibfnamefont{D.}~\bibnamefont{Prabhakaran}},
  \bibinfo{author}{\bibfnamefont{A.~T.} \bibnamefont{Boothroyd}},
  \bibinfo{author}{\bibfnamefont{P.~D.} \bibnamefont{Hatton}},
  \bibnamefont{and}
  \bibinfo{author}{\bibfnamefont{M.}~\bibnamefont{Altarelli}},
  \bibinfo{journal}{Phys.~Rev.~B} \textbf{\bibinfo{volume}{71}},
  \bibinfo{pages}{245102} (\bibinfo{year}{2005}).

\bibitem[{\citenamefont{Stoji\'c et~al.}(2005)\citenamefont{Stoji\'c, Binggeli,
  and Altarelli}}]{stojic:05a}
\bibinfo{author}{\bibfnamefont{N.}~\bibnamefont{Stoji\'c}},
  \bibinfo{author}{\bibfnamefont{N.}~\bibnamefont{Binggeli}}, \bibnamefont{and}
  \bibinfo{author}{\bibfnamefont{M.}~\bibnamefont{Altarelli}},
  \bibinfo{journal}{Phys.~Rev.~B} \textbf{\bibinfo{volume}{72}},
  \bibinfo{pages}{104108} (\bibinfo{year}{2005}).

\bibitem[{\citenamefont{Gibbs et~al.}(1988)\citenamefont{Gibbs, Harshman,
  Isaacs, McWhan, Mills, and Vettier}}]{gibbs:88a}
\bibinfo{author}{\bibfnamefont{D.}~\bibnamefont{Gibbs}},
  \bibinfo{author}{\bibfnamefont{D.~R.} \bibnamefont{Harshman}},
  \bibinfo{author}{\bibfnamefont{E.~D.} \bibnamefont{Isaacs}},
  \bibinfo{author}{\bibfnamefont{D.~B.} \bibnamefont{McWhan}},
  \bibinfo{author}{\bibfnamefont{D.}~\bibnamefont{Mills}}, \bibnamefont{and}
  \bibinfo{author}{\bibfnamefont{C.}~\bibnamefont{Vettier}},
  \bibinfo{journal}{Phys.~Rev.~Lett.} \textbf{\bibinfo{volume}{61}},
  \bibinfo{pages}{1241} (\bibinfo{year}{1988}).

\bibitem[{\citenamefont{Hannon et~al.}(1988)\citenamefont{Hannon, Trammell,
  Blume, and Gibbs}}]{hannon:88a}
\bibinfo{author}{\bibfnamefont{J.~P.} \bibnamefont{Hannon}},
  \bibinfo{author}{\bibfnamefont{G.~T.} \bibnamefont{Trammell}},
  \bibinfo{author}{\bibfnamefont{M.}~\bibnamefont{Blume}}, \bibnamefont{and}
  \bibinfo{author}{\bibfnamefont{D.}~\bibnamefont{Gibbs}},
  \bibinfo{journal}{Phys.~Rev.~Lett.} \textbf{\bibinfo{volume}{61}}
  \bibinfo{pages}{1245} (\bibinfo{year}{1988}).

\bibitem[{\citenamefont{Sch\"u{\ss}ler-Langeheine
  et~al.}(2001)\citenamefont{Sch\"u{\ss}ler-Langeheine, Weschke, Grigoriev,
  Ott, Meier, Vyalikh, Mazumdar, Sutter, Abernathy, Gr\"ubel
  et~al.}}]{schuessler:01a}
\bibinfo{author}{\bibfnamefont{C.}~\bibnamefont{Sch\"u{\ss}ler-Langeheine}},
  \bibinfo{author}{\bibfnamefont{E.}~\bibnamefont{Weschke}},
  \bibinfo{author}{\bibfnamefont{A.~Y.} \bibnamefont{Grigoriev}},
  \bibinfo{author}{\bibfnamefont{H.}~\bibnamefont{Ott}},
  \bibinfo{author}{\bibfnamefont{R.}~\bibnamefont{Meier}},
  \bibinfo{author}{\bibfnamefont{D.~V.} \bibnamefont{Vyalikh}},
  \bibinfo{author}{\bibfnamefont{C.}~\bibnamefont{Mazumdar}},
  \bibinfo{author}{\bibfnamefont{C.}~\bibnamefont{Sutter}},
  \bibinfo{author}{\bibfnamefont{D.}~\bibnamefont{Abernathy}},
  \bibinfo{author}{\bibfnamefont{G.}~\bibnamefont{Gr\"ubel}},
  \bibnamefont{et~al.}, \bibinfo{journal}{Journ. Electron. Spectrosc Relat.
  Phenom} \textbf{\bibinfo{volume}{114-116}}, \bibinfo{pages}{953}
  (\bibinfo{year}{2001}).

\bibitem[{\citenamefont{Ott et~al.}(2006)\citenamefont{Ott,
  Sch\"u{\ss}ler-Langeheine, Schierle, Grigoriev, Leiner, Zabel, Kaindl, and
  Weschke}}]{ott:06a}
\bibinfo{author}{\bibfnamefont{H.}~\bibnamefont{Ott}},
  \bibinfo{author}{\bibfnamefont{C.}~\bibnamefont{Sch\"u{\ss}ler-Langeheine}},
  \bibinfo{author}{\bibfnamefont{E.}~\bibnamefont{Schierle}},
  \bibinfo{author}{\bibfnamefont{A.~Y.} \bibnamefont{Grigoriev}},
  \bibinfo{author}{\bibfnamefont{V.}~\bibnamefont{Leiner}},
  \bibinfo{author}{\bibfnamefont{H.}~\bibnamefont{Zabel}},
  \bibinfo{author}{\bibfnamefont{G.}~\bibnamefont{Kaindl}}, \bibnamefont{and}
  \bibinfo{author}{\bibfnamefont{E.}~\bibnamefont{Weschke}},
  \bibinfo{journal}{Phys.~Rev.~B} \textbf{\bibinfo{volume}{74}},
  \bibinfo{pages}{094412} (\bibinfo{year}{2006}).

\bibitem[{\citenamefont{Koehler et~al.}(1967)\citenamefont{Koehler, Cable,
  Wilkinson, and Wollan}}]{koehler:67a}
\bibinfo{author}{\bibfnamefont{W.~C.} \bibnamefont{Koehler}},
  \bibinfo{author}{\bibfnamefont{J.~W.} \bibnamefont{Cable}},
  \bibinfo{author}{\bibfnamefont{M.~K.} \bibnamefont{Wilkinson}},
  \bibnamefont{and} \bibinfo{author}{\bibfnamefont{E.~O.}
  \bibnamefont{Wollan}}, \bibinfo{journal}{Phys. Rev.}
  \textbf{\bibinfo{volume}{151}}, \bibinfo{pages}{414} (\bibinfo{year}{1967}).

\bibitem[{\citenamefont{Spencer et~al.}(2005)\citenamefont{Spencer, Wilkins,
  Hatton, Brown, Hase, Purton, and Fort}}]{spencer:05b}
\bibinfo{author}{\bibfnamefont{P.~D.} \bibnamefont{Spencer}},
  \bibinfo{author}{\bibfnamefont{S.~B.} \bibnamefont{Wilkins}},
  \bibinfo{author}{\bibfnamefont{P.~D.} \bibnamefont{Hatton}},
  \bibinfo{author}{\bibfnamefont{S.~D.} \bibnamefont{Brown}},
  \bibinfo{author}{\bibfnamefont{T.~P.~A.} \bibnamefont{Hase}},
  \bibinfo{author}{\bibfnamefont{J.~A.} \bibnamefont{Purton}},
  \bibnamefont{and} \bibinfo{author}{\bibfnamefont{D.}~\bibnamefont{Fort}},
  \bibinfo{journal}{J. Phys. Cond. Mat.} \textbf{\bibinfo{volume}{17}},
  \bibinfo{pages}{1725} (\bibinfo{year}{2005}).

\bibitem[{\citenamefont{Leiner et~al.}(2000)\citenamefont{Leiner, Labergerie,
  Siebrecht, Sutter, and Zabel}}]{leiner:00a}
\bibinfo{author}{\bibfnamefont{V.}~\bibnamefont{Leiner}},
  \bibinfo{author}{\bibfnamefont{D.}~\bibnamefont{Labergerie}},
  \bibinfo{author}{\bibfnamefont{R.}~\bibnamefont{Siebrecht}},
  \bibinfo{author}{\bibfnamefont{C.}~\bibnamefont{Sutter}}, \bibnamefont{and}
  \bibinfo{author}{\bibfnamefont{H.}~\bibnamefont{Zabel}},
  \bibinfo{journal}{Physica B: Cond. Mat.} \textbf{\bibinfo{volume}{283}},
  \bibinfo{pages}{167} (\bibinfo{year}{2000}).

\bibitem[{\citenamefont{Weschke et~al.}(2004)\citenamefont{Weschke, Ott,
  Schierle, Sch\"u{\ss}ler-Langeheine, Vyalikh, Kaindl, Leiner, Ay, Schmitte,
  Zabel et~al.}}]{weschke:04a}
\bibinfo{author}{\bibfnamefont{E.}~\bibnamefont{Weschke}},
  \bibinfo{author}{\bibfnamefont{H.}~\bibnamefont{Ott}},
  \bibinfo{author}{\bibfnamefont{E.}~\bibnamefont{Schierle}},
  \bibinfo{author}{\bibfnamefont{C.}~\bibnamefont{Sch\"u{\ss}ler-Langeheine}},
  \bibinfo{author}{\bibfnamefont{D.~V.} \bibnamefont{Vyalikh}},
  \bibinfo{author}{\bibfnamefont{G.}~\bibnamefont{Kaindl}},
  \bibinfo{author}{\bibfnamefont{V.}~\bibnamefont{Leiner}},
  \bibinfo{author}{\bibfnamefont{M.}~\bibnamefont{Ay}},
  \bibinfo{author}{\bibfnamefont{T.}~\bibnamefont{Schmitte}},
  \bibinfo{author}{\bibfnamefont{H.}~\bibnamefont{Zabel}},
  \bibnamefont{et~al.}, \bibinfo{journal}{Phys.~Rev.~Lett.}
  \textbf{\bibinfo{volume}{93}}, \bibinfo{pages}{157204}
  (\bibinfo{year}{2004}).

\bibitem[{\citenamefont{Thole et~al.}(1985)\citenamefont{Thole, van~der Laan,
  Fuggle, Sawatzky, Karnatak, and Esteva}}]{Thole85}
\bibinfo{author}{\bibfnamefont{B.~T.} \bibnamefont{Thole}},
  \bibinfo{author}{\bibfnamefont{G.}~\bibnamefont{van~der Laan}},
  \bibinfo{author}{\bibfnamefont{J.~C.} \bibnamefont{Fuggle}},
  \bibinfo{author}{\bibfnamefont{G.~A.} \bibnamefont{Sawatzky}},
  \bibinfo{author}{\bibfnamefont{R.~C.} \bibnamefont{Karnatak}},
  \bibnamefont{and} \bibinfo{author}{\bibfnamefont{J.-M.}
  \bibnamefont{Esteva}}, \bibinfo{journal}{Phys. Rev. B}
  \textbf{\bibinfo{volume}{32}}, \bibinfo{pages}{5107} (\bibinfo{year}{1985}).

\bibitem[{\citenamefont{Cowan}(1981)}]{Cowan81}
\bibinfo{author}{\bibfnamefont{R.~D.} \bibnamefont{Cowan}},
  \emph{\bibinfo{title}{The theory of Atomic structure and spectra}}
  (\bibinfo{publisher}{California press}, \bibinfo{year}{1981}).

\bibitem[{\citenamefont{Newton}(1976)}]{Newton76}
\bibinfo{author}{\bibfnamefont{R.~G.} \bibnamefont{Newton}},
  \bibinfo{journal}{Am. J. Phys.} \textbf{\bibinfo{volume}{44}},
  \bibinfo{pages}{639} (\bibinfo{year}{1976}).

\end{thebibliography}
\end{document}